\documentclass[traditabstract]{aa} 

\usepackage[pdftex,final]{graphicx}  
\usepackage[comma,authoryear]{natbib} 
\bibpunct{(}{)}{,}{a}{}{,}            

\begin{document}

   \title{Sensitivity of ring analysis to near-surface magnetism}

   \subtitle{}

   \author{T. Corbard
          }

   \institute{Universit\'e de Nice Sophia-Antipolis, CNRS, Observatoire de la C\^ote d'Azur,
             BP 4229, 06304 Nice Cedex 4, France\\
              \email{Thierry.Corbard@oca.eu}
             }

  \abstract
   {While magnetic field has only second order effects on global sound waves, it 
   can affect the mode frequencies and flow fields inferred using local helioseismology.
   A strong localized field can bias our results when one try to infer global scale flows 
   such as the meridional circulation. On the other hand, 
   one important object of research in this area is to determine whether one can detect
   changes in frequencies or sub-surface flow fields that could be a precursor of 
   surface magnetic events. In this paper we review recent development in this area focusing on results obtained 
   with the ring-diagram analysis technique. 
    }


   \maketitle

\section{Introduction}
When using global helioseismology to infer internal flows, we can sense only the part of the rotation that is symmetric about the equator and
sensitivity kernels are independent of longitude which prevent us  from studying for instance the interaction between an active region and sub-surface flows. It is moreover impossible to separate the spherically asymmetric effects other than rotation (meridional circulation, magnetic fields, structural asphericity) and  it is difficult to fit high
degree modes in global analysis due to mode leakage.
These difficulties can be partially overcome with local helioseismology:  because the wavelength of high order modes is small compared with the typical scale over which equilibrium structure changes, the modes can be approximated locally by plane sound waves. High degree acoustic waves are damped and cannot travel around the full circumference of the sun and their frequencies are local measures of the sun's properties. 
In ring diagram analysis \citep{1988ApJ...333..996H}, we typically analyze a mosaic of 16 degree patches tracked over about one day on Doppler
 images taken at 1 mn cadence. Peaks fitted in the 3D power spectra are typically in the frequency range from 1.7 mHz to 5.6 mHZ and include the signature of modes with very short horizontal wavelengths corresponding to spherical harmonic degrees ranging from $\ell=200$ to $\ell=1100$ and short lifetimes.
Local effects are thus expected to be more important for these modes. Some attempts have been made to detect magnetic effect directly by tracking the expected signature in the spectra when considering the contribution of the Alfv\'en velocity in the dispersion relation but this a is second order effect only and this research has not been conclusive to date \citep{1996AAS...188.6905H}.  The ring shape of slices through the 3D power spectra are therefore mainly affected by the advection induced by sub-photospheric flows in both meridional en azimuthal directions but the presence of magnetic field in the tracked area also affect amplitude, power and frequencies of the peaks. This kind of analysis also allows us to make statistics about many fluid descriptors such as  amplitude, direction, helicity or vorticity as a function of the average magnetic field amplitude within the patch \citep{2004ApJ...605..554K,2005ApJ...631..636K} .


\section{Effect of magnetic field on high degree modes sensed by rings }
We know that for instance sunspots scatter and absorb p-modes. More generally, amplitude of solar oscillations decrease significantly in active regions either due to the  absorption of acoustic modes or to a  weaker excitation in active regions.
There are well known temporal variations of global modes mean frequencies, amplitude and linewidths with the cycle which are likely linked to global magnetic field variations. With ring analysis, we can isolate active regions where the magnetic field amplitude is much higher than the mean global field and therefore we expect
these effects to be magnified.
In \citet{2001ApJ...563..410R} pairs of active and quiet regions located at the same latitude have been analyzed showing a monotonic increase of mode frequencies with increasing magnetic activity. In the five minute band, a decrease in both mode power and lifetimes is clearly detected while,  at higher frequencies, \citet{2004ApJ...608..562H} have shown that the trend is reversed 
 revealing a regime around 5 mHz where the modes are absorbed like p-modes in presence of strong magnetic fields and enhanced like high-frequency waves at lower activity levels. 
When looking at the source location of coronal mass ejections (CMEs) with low value of the magnetic flux,
\citet{2008JApA...29..207T} realized that the oscillation modes have higher life times than other quiet regions thus indicating a slower damping process.  They pointed out that this information may be use 
to forecast active region which may trigger CMEs.
   

   \begin{figure}[htbp]
     \begin{center}
       \includegraphics[width=8cm]{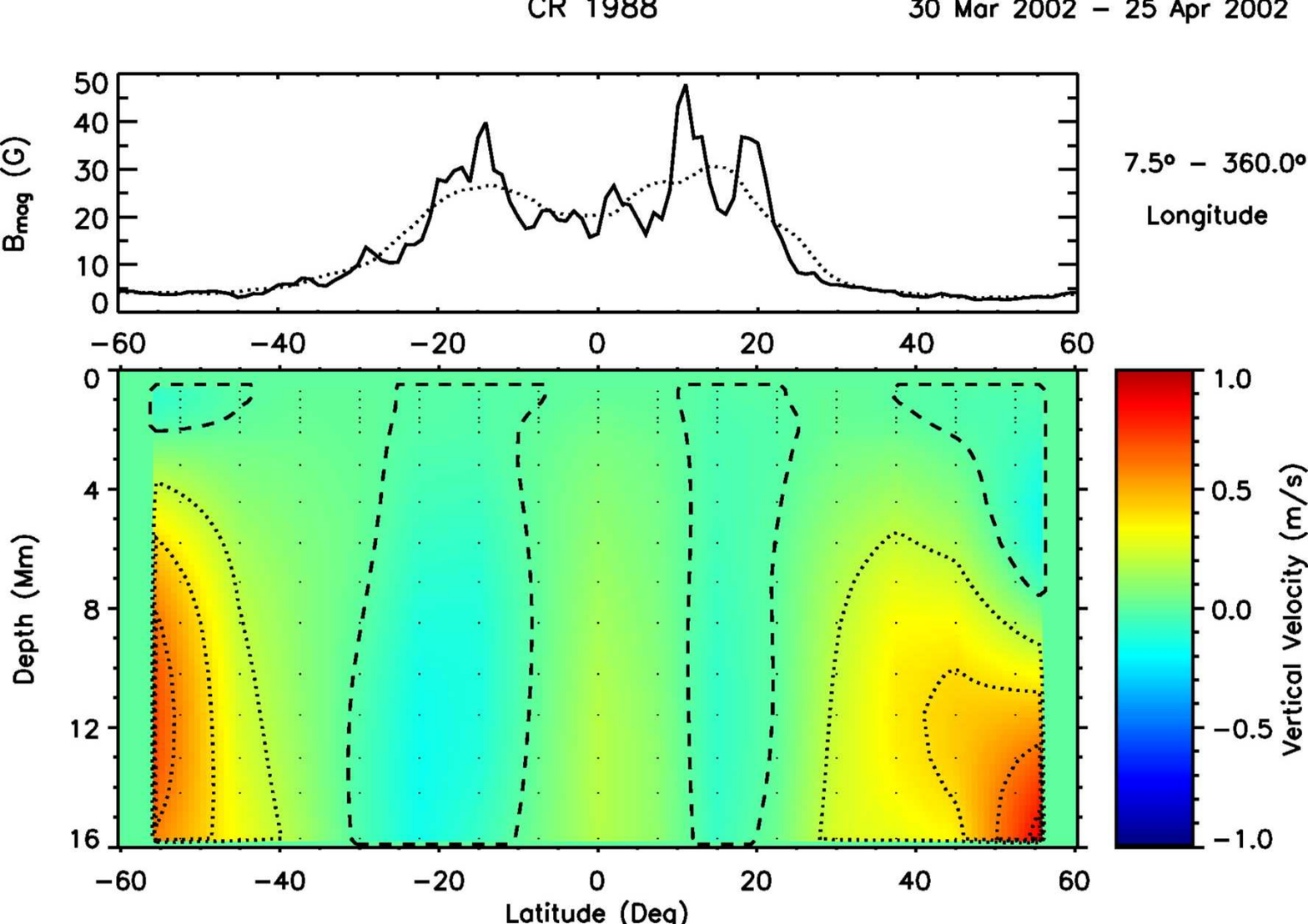}
     \end{center}
     \caption{Vertical velocity component averaged over Carrington rotation CR 1988 (2002 March 30April 25) as a function of latitude and depth. Top: Surface magnetic flux as a function of latitude (solid line) and averaged over 15° (dotted curve). Bottom: Vertical velocity derived from GONG data after removing the large-scale flow components.  \citep{2004ApJ...605..554K}}
     \label{fig:komm}
   \end{figure}

\section{Effect of magnetic field on subsurface flows sensed by rings}
The depth dependence of the horizontal velocity field can be inverted from theoretical sensitivity 
kernels which are essentially given by the mode kinetic energy density \citep{2007ApJ...662..730B}. 
These kernels however are symmetric and cancel out over the analysis patch for vertical flows. 
One way to get hints 
on the vertical flows is to assume mass conservation and to compute it from the divergence
of the horizontal components \citep{2004ApJ...605..554K}. In general low surface magnetic activity 
corresponds to sub-surface upflows while medium magnetic activity is more likely associated with downflows. Figure \ref{fig:komm} shows, for one Carrington rotation, the longitudinal average of the vertical velocities down to 16 Mm. Weak downflows of about 1 ms$^{-1}$ are detected near active latitudes. 
But for strong magnetic field associated with surface magnetic features we can detect downflows down to about
10 Mm where there is a transition to upflows which are also typical for flux ropes in convectively
 unstable layers.
 
 Concerning the meridional flows, most analysis agree on a poleward flow of about 10 to 20 meters 
 per second increasing with depth and a secondary flow that converge toward the mean latitude of activity
 \citep{2002ApJ...570..855H,2004ApJ...603..776Z,2005ApJ...631..636K}.
 There are strong debate on the existence of counter cells at high latitudes \citep{2002ApJ...570..855H} where geometric calibration 
 issues enter in consideration in a critical way \citep{2006ApJ...638..576G}. While studying how surface magnetic 
activity influence our inference on meridional flows, \citet{2008SoPh..252..235G} found that inflows associated to activity are confined to the upper
layers  but persist even after aggressively masking surface activity. It has also been shown by \citet{2006SoPh..236..227Z} using ring diagrams that the north-south asymmetry of the flow reflects the north-south asymmetry of the magnetic flux.

One interesting fluid descriptor that can be derived from horizontal flow maps is the kinetic helicity. Large helicity might be associated to highly twisted magnetic flux tubes that are likely to produce flare. \citet{2005ApJ...630.1184K} reported effectively a high correlation between the maximum kinetic helicity density and X-flare. From an analysis of 
13 emerging regions, they investigate the question whether the increase of kinetic helicity coincides with  the emergence of flux and obtained a first result showing that the kinetic helicity in shallow layers lags behind 
the kinetic helicity in deeper layers which, if proved to be statistically significant, could be used in the future
to predict the emergence of new flux and flare \citep{2008ApJ...672.1254K}.

\section{Conclusions}
Ring analysis is sensitive to activity (to local and maybe also global field).
Sensitivity to local field can be seen as a source of noise when one try to infer global flows 
(meridional, zonal) and how they vary with the cycle. An important question is then how to
 filter these effects and to what level.
 Pattern of global flows are mostly independent of activity level but their magnitude vary.
When patches are sorted as a function of some magnetic activity index, it gives some hints on how
 the flow are organized below and around active regions.  
Lots of efforts are now made in trying to detect correlations between activity events (CMEs, flares, filaments)
 and some descriptors of the submerge flows (helicity) or underlying waves (linewidth) essentially in the hope of
  finding precursor indicators.
  
A lot of efforts have been made during the last few years that should be continued in
making comparison between the different methods,
understanding errors, methods resolution and correlations,
understanding effects of image misalignment and  geometric effects.

Finally we should mention that, together with the upcoming SDO mission
 which will provide higher resolution data, 
there are  a lot of new developments in this field such as 
high resolution ring analysis using f-modes \citep{2006ApJ...653..725H},
deeply penetrating ring techniques \citep{2006ApJ...638..576G},
full 3D inversion procedures or structural inversions for ring frequency shifts \citep{2004ApJ...610.1157B}.

\begin{acknowledgements}
Participation of GW at the HELAS NA3-4 workshop was supported by the European
Helio- and Asteroseismology Network (HELAS), a major international collaboration
funded by the European Commission's Sixth Framework Programme.
\end{acknowledgements}

\bibliographystyle{aa}
\bibliography{biblio}
\end{document}